\documentclass[pra,twocolumn,groupedaddress]{revtex4-1}

\usepackage{graphicx}
\usepackage{color}
\begin{document}


\title{Investigating Energy Scales of Fractional Quantum Hall States \\ using Scanning Gate Microscopy}


\author{B. A. Braem}
\email[]{bbraem@phys.ethz.ch}
\affiliation{ETH Z\"urich, Solid State Physics Laboratory, Otto-Stern-Weg 1, 8093 Z\"urich, Switzerland}

\author{T. Kr\"ahenmann}
\affiliation{ETH Z\"urich, Solid State Physics Laboratory, Otto-Stern-Weg 1, 8093 Z\"urich, Switzerland}
\author{S. Hennel}
\affiliation{ETH Z\"urich, Solid State Physics Laboratory, Otto-Stern-Weg 1, 8093 Z\"urich, Switzerland}
\author{C. Reichl}
\affiliation{ETH Z\"urich, Solid State Physics Laboratory, Otto-Stern-Weg 1, 8093 Z\"urich, Switzerland}
\author{W. Wegscheider}
\affiliation{ETH Z\"urich, Solid State Physics Laboratory, Otto-Stern-Weg 1, 8093 Z\"urich, Switzerland}
\author{K. Ensslin}
\affiliation{ETH Z\"urich, Solid State Physics Laboratory, Otto-Stern-Weg 1, 8093 Z\"urich, Switzerland}
\author{T. Ihn}
\affiliation{ETH Z\"urich, Solid State Physics Laboratory, Otto-Stern-Weg 1, 8093 Z\"urich, Switzerland}


\date{\today}

\begin{abstract}
We use the voltage biased tip of a scanning force microscope at a temperature of 35\,mK to locally induce the fractional quantum Hall state of $\nu=1/3$ in a split-gate defined constriction. Different tip positions allow us to vary the potential landscape. From temperature dependence of the conductance plateau at $G=1/3 \times e^2/h$ we determine the activation energy of this local $\nu=1/3$ state. We find that at a magnetic field of 6\,T the activation energy is between 153\,$\mu$eV and 194\,$\mu$eV independent of the shape of the confining potential, but about 50\% lower than for bulk samples.
\end{abstract}


\maketitle

\section{Introduction}
At high magnetic fields, two-dimensional electron systems display the integer 
\cite{klitzing_new_1980} and the fractional \cite{tsui_two-dimensional_1982} 
quantum Hall (QH) effect. In this regime the current is expected to be carried 
by counterpropagating chiral edge channels which interact only if they are 
brought in proximity by a quantum point contact (QPC). 
This interaction could be exploited to check for non-Abelian statistics using a 
Fabry-Perot interferometer in the fractional QH regime  \cite{de_c._chamon_two_1997}. Even though fragile states such as $\nu=5/2$ have been observed in single constrictions \cite{radu_quasi-particle_2008, lin_measurements_2012, baer_experimental_2014}, only interferometers in the regime of integer and more robust fractional QH states have been shown \cite{camino_realization_2005, camino_$e/3$_2007, kou_coulomb_2012, willett_magnetic-field-tuned_2013}. Therefore a deeper understanding of fractional QH states in constrictions is required.

For this purpose we form different potential landscapes and measure the 
activation energy of the fractional QH state of $\nu=1/3$ confined between edge 
channels. We use a movable "scanning" gate to tune the potential of a 
split-gate defined QPC.

For scanning gate microscopy (SGM) we apply a voltage to the conductive tip of a 
scanning force microscope to form the movable gate. This technique gives an 
additional spatial degree of freedom compared to conventional top gate-defined 
structures. SGM has been successfully used to obtain spatial information about 
integer and fractional QH edge channels \cite{aoki_imaging_2005, paradiso_imaging_2012, pascher_imaging_2014}. The activation energy of the fractional QH state $\nu=1/3$ has been found to be lower for confined geometries than for bulk samples \cite{baer_interplay_2014, dethlefsen_signatures_2006}. Our measurements confirm this finding and indicate that the activation energy of a confined fractional QH state depends only weakly on the exact shape of the confining potential.

\section{Experimental details}
The experiment is performed in a home-built scanning force microscope in a dilution refrigerator with a base temperature of 35\,mK  \cite{gildemeister_construction_2007}. Coulomb blockade measurements (data not shown) in this setup showed an electron temperature of $\approx 200$\,mK. The scanning force sensor is a focused-ion-beam sharpened PtIr-wire attached to a quartz tuning fork \cite{rychen_low-temperature_1999}. To locate the sample, we use the tip for topography scanning. During SGM measurements, the tip is retracted and scanned at a constant distance of 110\,nm above the sample surface. It is only capacitively coupled to the electronic system of the sample  and no current flows from the tip to the sample. The direction of fast tip motion is referred to as the $x$-direction.

\begin{figure}
 \includegraphics[width=\linewidth]{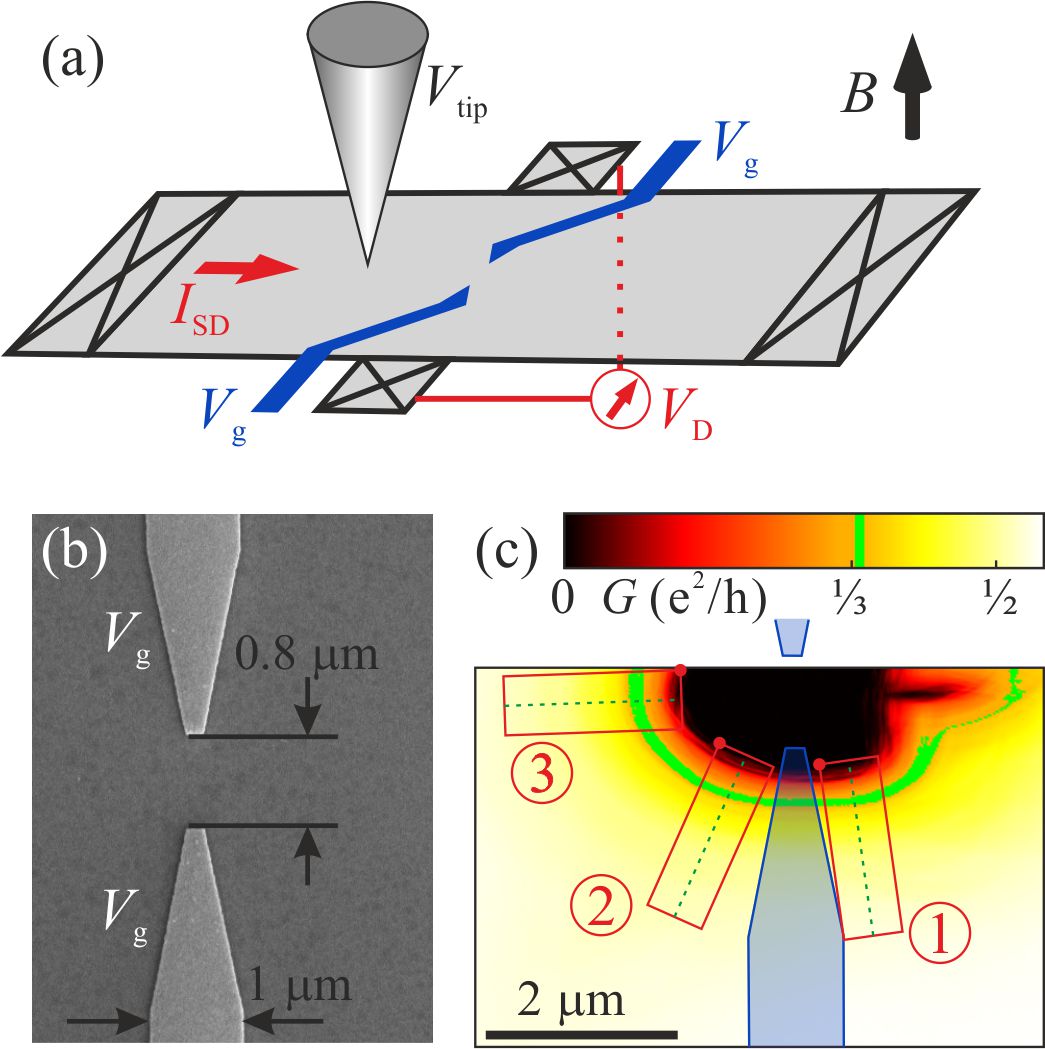}
 \caption{(a) Schematic of the setup with the Hall bar (grey) and the QPC defined by top gates (blue). The tip is biased with -10\,V and kept at a fixed distance of 110\,nm to the GaAs surface. (b) Scanning electron microscope image of the sample showing the top gates (light gray) that define the constriction. (c) Conductance $G(x,y)$ as function of tip position. The position of the top gates is marked in blue and red rectangles indicate the high-resolution scan frames described later.}
 \label{fig1label}
\end{figure}

The sample is a 400\,$\mu$m wide Hall bar etched into a GaAs-AlGaAs heterostructure with a 310\,nm deep two-dimensional electron gas (2DEG). The 2DEG has an electron density of $9.3\times10^{10}\mathrm{ cm}^{-2}$ and a mobility $4.7\times 10^6 \mathrm{ cm}^2\mathrm{/Vs}$ at 35\,mK. The nanostructure is a split-gate defined quantum point contact with a lithographic width of 800\,nm as shown in the scanning electron microscope image in Fig. \ref{fig1label}(b). 

Figure \ref{fig1label}(a) is a schematic of the measurement setup. The constriction within the 2DEG is defined by applying a gate voltage $V_{\mathrm{g}}=-0.7$\,V to the top gates and a local perturbation is created by the dc tip bias $V_\mathrm{tip}=-10$\,V. 
We determine the four-terminal conductance $G=I_{\mathrm{SD}}/V_\mathrm{D}$ by applying a voltage $V_{\mathrm{SD}}$ between source and drain of the Hall bar and measuring the source-drain current $I_{\mathrm{SD}}$ and the diagonal voltage $V_{\mathrm{D}}$. For lock-in measurements an ac excitation is applied, therefore $V_{\mathrm{SD}}=V_{\mathrm{SD-dc}}+V_{\mathrm{SD-ac}}$. If not stated explicitly, $V_{\mathrm{SD-dc}}=0$ and $V_{\mathrm{SD-ac}}=20\,\mu$V at 68\,Hz.

A magnetic field $B=6$\,T is applied normal to the plane of the 2DEG corresponding to a quantum Hall filling factor $\nu_{\mathrm{bulk}}=2/3$. We determine the effective filling factor $\nu_{\mathrm{QPC}}$ in the constriction from the diagonal conductance $G=I_{\mathrm{SD}}/V_{\mathrm{D}}=\nu_{\mathrm{QPC}}\,e^2/h$ \cite{beenakker_quantum_1991, pascher_imaging_2014}.

\section{Spatial mapping of locally induced fractional filling factors}
Figure \ref{fig1label}(c) shows the conductance $G(x,y)$ as a function of tip position $(x,y)$ \footnote{The tip can not be moved over the upper top gate because of an obstacle that would cause a tip crash.}. Without the tip induced perturbation, the potential landscape of the constriction can be approximated by a saddle point \cite{buettiker_quantized_1990}. Bringing the tip closer adds two additional constrictions, one between each gate and the region of depleted 2DEG below the tip \cite{pascher_resonant_2014-1}. The tip position determines which of the three constrictions will limit the diagonal conductance by having the lowest electron density and thus forming the lowest filling factor $\nu_{\mathrm{min}}$ on the path from source to drain.

To study different confining potentials we chose three $1.5\,\mu\textrm{m} \times 0.5 \,\mu\textrm{m}$ high resolution scan frames \textcircled{\raisebox{-1pt}{1}} - \textcircled{\raisebox{-1pt}{3}}. Their positions are marked by red rectangles in Fig. \ref{fig1label}(c). Dots indicate the origins of their respective coordinate systems ($x$-axis parallel to long side). Scan frames \textcircled{\raisebox{-1pt}{1}} and \textcircled{\raisebox{-1pt}{2}} are chosen to slightly modify and shift the split-gate defined constriction. \textcircled{\raisebox{-1pt}{3}} is placed along the direction of current flow through the QPC such that the constriction of lowest filling factor $\nu_{\mathrm{min}}$ is shifted towards the left hand side. 

\begin{figure}
 \includegraphics[width=\linewidth]{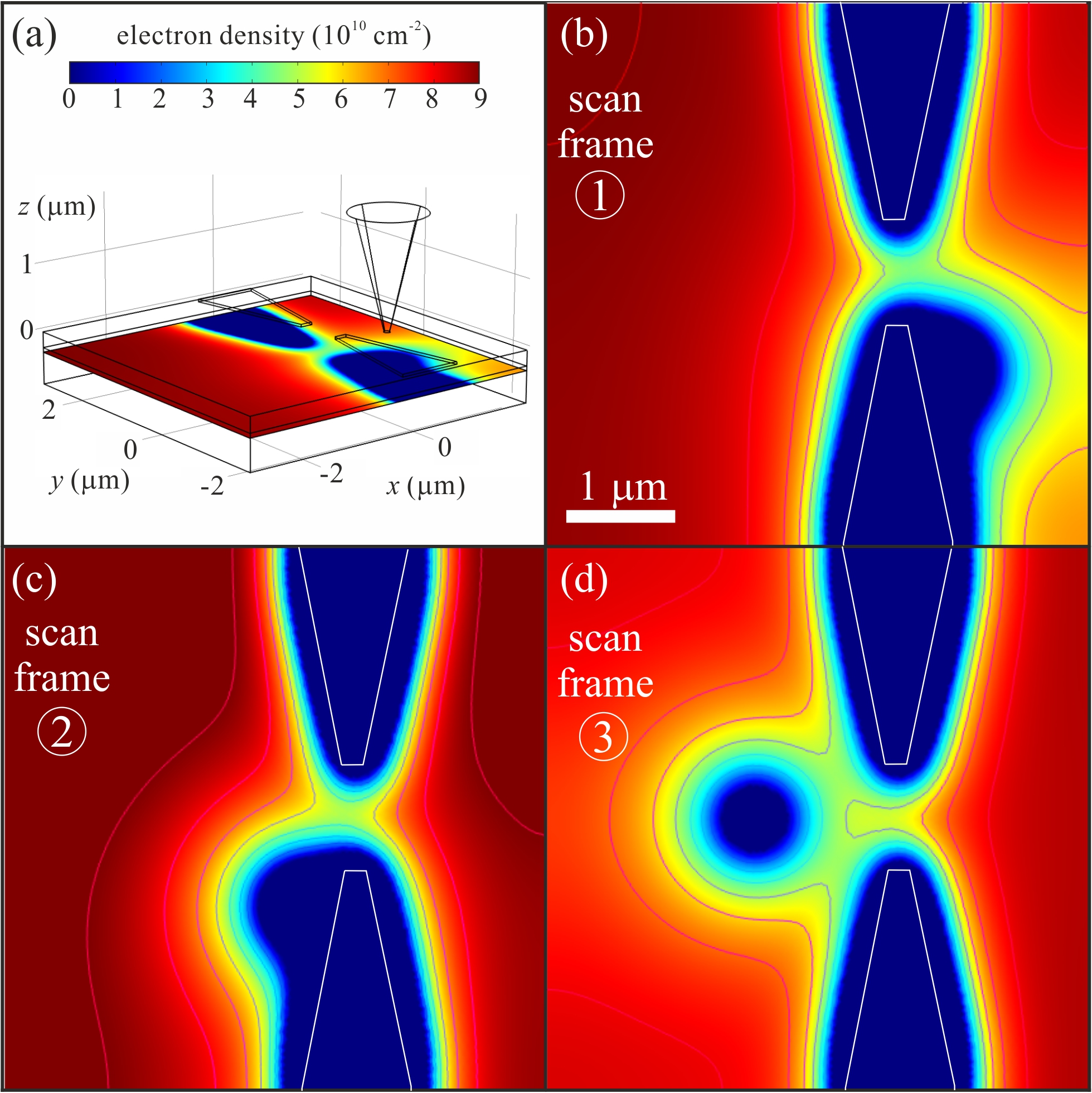}
 \caption{Electron density calculated by finite element simulation using Thomas-Fermi approximation: (a) Model geometry with split-gates and tip, electron density in color scale. (b)-(d) Electron density calculated for tip positions in scan frames \textcircled{\protect\raisebox{-1pt}{1}}, \textcircled{\protect\raisebox{-1pt}{2}}, and \textcircled{\protect\raisebox{-1pt}{3}} show the shift of the constriction.}
 \label{fig2label}
\end{figure}

We investigate the effect of the different tip positions by calculating the electron density in the vicinity of the constriction using finite-element simulation \footnote{Electrostatic study using \emph{COMSOL 5.0}}. The model includes Thomas-Fermi screening \cite{ando_electronic_1982} and finite thickness of the 2DEG \cite{ando_electronic_1982, fang_negative_1966}. For each scan frame we use the tip coordinate at the intersection of the green dashed line and the $\nu=1/3$ stripe in Fig. \ref{fig1label}(c) and a tip-surface distance of 110\,nm. Figure \ref{fig2label}(a) illustrates the model geometry for scan frame \textcircled{\raisebox{-1pt}{1}} and the electron density of the 2DEG in color scale. The calculated electron densities for scan frames \textcircled{\raisebox{-1pt}{1}} - \textcircled{\raisebox{-1pt}{3}} in Figs. \ref{fig2label}(b)-(d) show that the constriction with lowest electron density $\nu_{\mathrm{min}}$ lies inside the lithographic gap for \textcircled{\raisebox{-1pt}{1}} and \textcircled{\raisebox{-1pt}{2}}, but is shifted by $\approx 500\, \textrm{nm}$ in \textcircled{\raisebox{-1pt}{3}}. Therefore the potential landscape at $\nu_{\mathrm{min}}$ is not only modified by the tip potential, but also by the lateral shift with respect to the disorder potential.

We measured $G(x,y)$ in \textcircled{\raisebox{-1pt}{1}}, \textcircled{\raisebox{-1pt}{2}}, and \textcircled{\raisebox{-1pt}{3}} with a horizontal resolution of 5\,nm at eight different mixing chamber temperatures $T$ between 35\,mK and 800\,mK. Results and evaluation will be discussed exemplarily for scan frame \textcircled{\raisebox{-1pt}{1}}, results from \textcircled{\raisebox{-1pt}{2}} and \textcircled{\raisebox{-1pt}{3}} have been analyzed in the same way.

\begin{figure}
 \includegraphics[width=\linewidth]{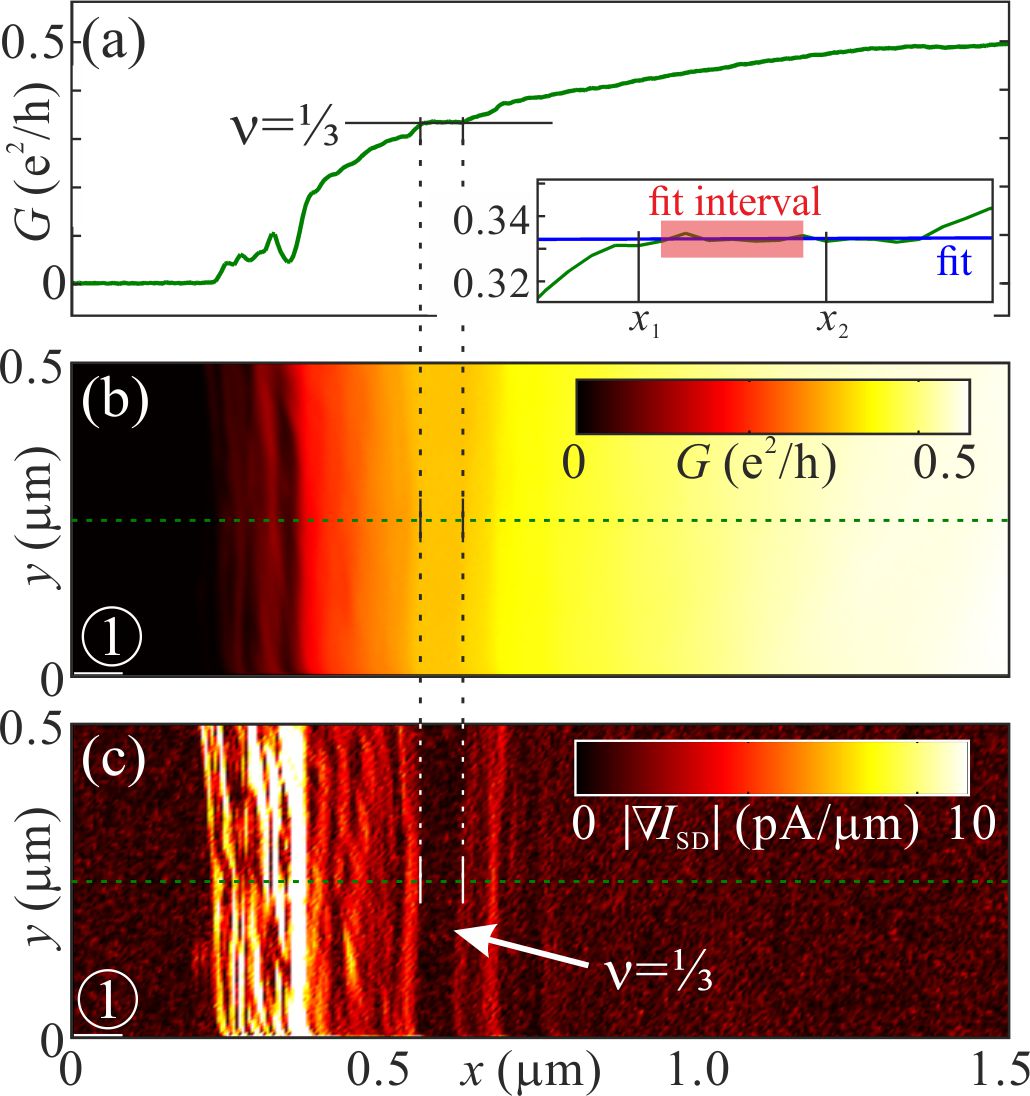}
 \caption{$G(x,y)$ in scan frame \textcircled{\protect\raisebox{-1pt}{1}}. (a) Cut $G(x,y=250\mathrm{\,nm})$ showing a pronounced $\nu=1/3$ plateau at $G\approx0.33\mathrm{\,e^2/h}$. Inset: Zoom to the plateau with linear fit in blue. (b) $G(x,y)$ and (c) $|\nabla I_{\mathrm{SD}}(x,y)|$ of scan frame \textcircled{\protect\raisebox{-1pt}{1}}. The modulus of the gradient of $I_{\mathrm{SD}}(x,y)$ shows the spatial extent of the $\nu=1/3$ plateau.}
 \label{fig3label}
\end{figure}

Figure \ref{fig3label}(a) shows $G(x,y=250\textrm{ nm})$ along the center line $y=250$\,nm of \textcircled{\raisebox{-1pt}{1}} at 35\,mK. The full data set $G(x,y)$ is displayed in Fig. \ref{fig3label}(b) as a color plot. The data shows a pronounced $\nu_{\mathrm{QPC}}=1/3$ plateau at $G=1/3 \times e^2/h$. The spatial extent of this plateau is visible as a dark belt in the numerically calculated modulus of the gradient of $I_\mathrm{SD}(x,y)$ \footnote{$ | \nabla I_{\mathrm{SD}}|=\sqrt{(\mathrm{d}I_{\mathrm{SD}}/\mathrm{d}x)^2+(\mathrm{d}I_{\mathrm{SD}}/\mathrm{d}y)^2}$} in Fig. \ref{fig3label}(c).
Repeating the measurement at higher temperature qualitatively shows the same features, but the plateau of $\nu_{\mathrm{QPC}}=1/3$ has an increased slope due to thermal activation. 

\section{Activation energies}
\begin{figure}
 \includegraphics[width=0.9\linewidth]{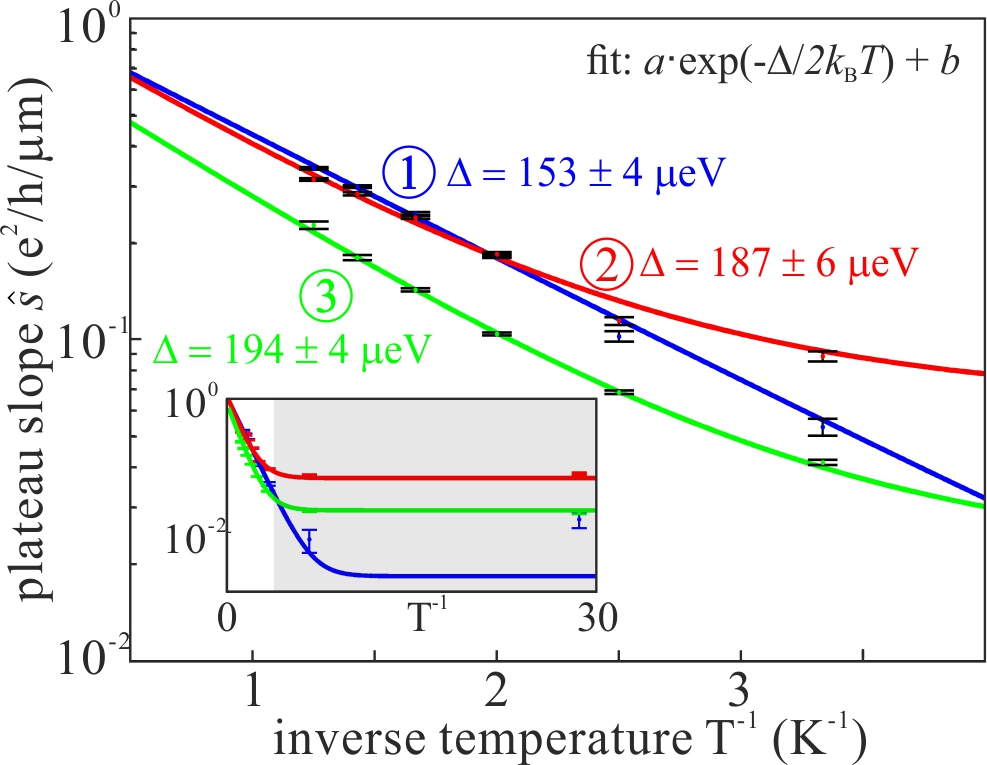}
 \caption{Arrhenius plot showing the mean slope of the $\nu=1/3$ plateau as function of temperature for all three scan frames. Fits $a\cdot \mathrm{exp(}-\Delta/2k_BT\mathrm{)}+b$ to the data give activation energies of $153 - 194 \,\mu$eV. Inset: entire temperature range showing the saturation at $T < 200$\,mK.}
 \label{fig4label}
\end{figure}

The temperature dependence of the slope of the $\nu_{\mathrm{QPC}}=1/3$ plateau is used to determine the activation energy $\Delta$ of the $\nu=1/3$ state \cite{wei_temperature_1985, baer_interplay_2014} in this confined geometry. The algorithm for determining the slope at each temperature $T$ by linear fitting is the following:

The onset $x_1(y)$ of the plateau of $G(x,y)$ for every scanned line $y$ is defined by the minimum of the smoothed d$^2G(x,y)/$d$^2x$. Analogously, the end of the plateau $x_2(y)$ is defined by the maximum of the curvature. The slope $s(y)$ of the plateau for every scan line $y$ is obtained by linear fitting $G(x,y)$ in the interval $[x_1+(x_2-x_1)/8, x_2-(x_2-x_1)/8]$. We calculate the mean slope $\hat{s}=\langle s(y) \rangle$ and repeat this algorithm for every scan frame \textcircled{\raisebox{-1pt}{1}} - \textcircled{\raisebox{-1pt}{3}} and all temperatures $T$. Figure \ref{fig4label} is an Arrhenius plot of the mean values $\hat{s}_\textrm{\textcircled{\raisebox{-1pt}{1}}}(T)$, $\hat{s}_\textrm{\textcircled{\raisebox{-1pt}{2}}}(T)$, and $\hat{s}_\textrm{\textcircled{\raisebox{-1pt}{3}}}(T)$. The uncertainties of the data points are given by the standard deviation of $s(y)$ for every scan frame and temperature. 

The mean plateau slopes $\hat{s}_\textrm{\textcircled{\raisebox{-1pt}{1}}}(T)$, $\hat{s}_\textrm{\textcircled{\raisebox{-1pt}{2}}}(T)$, and $\hat{s}_\textrm{\textcircled{\raisebox{-1pt}{3}}}(T)$ are fitted with an exponential temperature dependence $a\cdot \mathrm{exp(}-\Delta/2k_BT\mathrm{)}+b$. For the three scan frames we obtain activation energies $\Delta_\textrm{\textcircled{\raisebox{-1pt}{1}}}=153 \pm4\,\mu$eV,  $\Delta_\textrm{\textcircled{\raisebox{-1pt}{2}}}=187 \pm6 \,\mu$eV, and  $\Delta_\textrm{\textcircled{\raisebox{-1pt}{3}}}=194 \pm4 \,\mu$eV. The uncertainty of $\Delta$ is given by the square root of the corresponding diagonal element of the fit covariance matrix, but underestimates systematic errors in the measurement. Due to insufficient thermal anchoring of the sample leads, the electron temperature of the sample is higher than the cryostat temperature $T$. This leads to a non-zero saturation parameter $b$.

\section{Finite bias measurements}
Finite-bias measurements are an alternative way to measure the energy gap of quantum Hall states \cite{ebert_two-dimensional_1983, pascher_imaging_2014}. Increasing the source-drain bias voltage decreases the extent of QH plateaus until they vanish at a critical bias voltage $V_{\mathrm{crit}}$. In a single-particle picture, the energy $eV_{\mathrm{crit}}$ corresponds to the energy gap of the system \cite{berggren_characterization_1988}.

\begin{figure}
 \includegraphics[width=\linewidth]{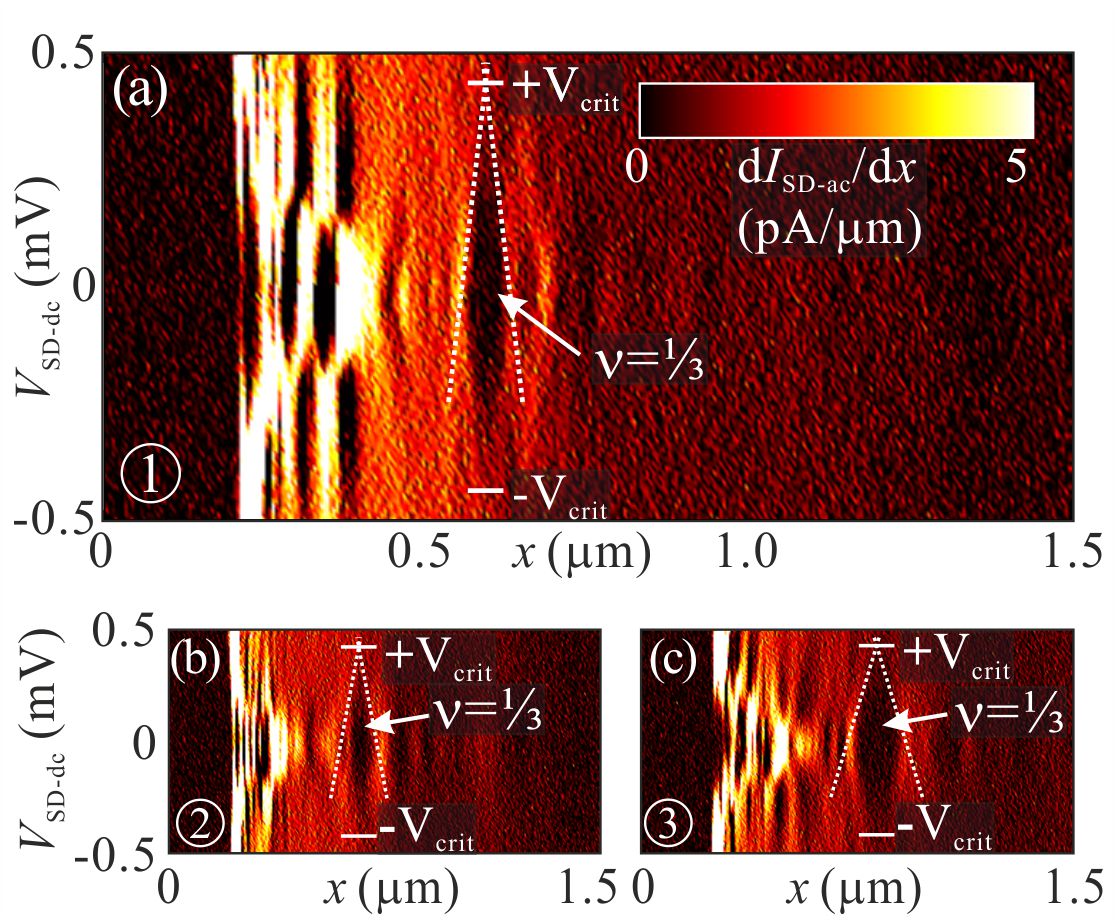}
 \caption{
 Numerical derivative of ac source drain current $\textrm{d}I_{\mathrm{SD-ac}}/\textrm{d}x$ at 35\,mK as function of dc bias voltage and tip position. (a) Scan frame \textcircled{\protect\raisebox{-1pt}{1}}, (b) scan frame \textcircled{\protect\raisebox{-1pt}{2}}, and (c) scan frame \textcircled{\protect\raisebox{-1pt}{3}}. The scanned lines are the center lines $y=250$\,nm of the high-resolution scan frames marked by green dashed lines in Fig. \ref{fig1label}(c). The $\nu=1/3$ finite bias diamonds in all scan frames end at $V_{\mathrm{crit}}\approx 400 \, \mu$V. Dotted lines are guide to the eye.
 }
 \label{fig5label}
\end{figure}

Additionally to the ac source-drain voltage $V_{\mathrm{SD-ac}}$ we applied a dc voltage $V_{\mathrm{SD-dc}}=[-0.5,+0.5]\,\mathrm{mV}$ and measured the conductance $G(x,V_{\mathrm{SD-dc}})$ at base temperature. The tip positions $x$ are on the center lines $y=250$\,nm of the scan frames (green dashed lines in Fig. \ref{fig1label}(c)). The numerical derivative of the ac source drain current $\textrm{d}I_{\mathrm{SD-ac}}/\textrm{d}x$ as function of V$_{\mathrm{SD-dc}}$ and tip position $x$ is shown in Fig. \ref{fig5label}. In the center of the images, the finite bias diamond of the $\nu=1/3$ state is visible. As indicated by the dotted lines, the diamonds of all scan frames close at a dc voltage \mbox{$V_{\mathrm{crit}}\approx 400 \,\mu$V}. In agreement with scanning gate measurements in the integer QH regime we find that $V_{\mathrm{crit}}$ is independent of tip position \cite{pascher_imaging_2014}.

\section{Conclusions}
We have used the flexibility of SGM to induce the fractional quantum Hall state $\nu=1/3$ in different confining potentials. The activation energies measured in scan frames with lateral tip position (\textcircled{\raisebox{-1pt}{1}} and \textcircled{\raisebox{-1pt}{2}}, which are expected to induce similar potential landscapes) differ by a value larger than the difference to the one measured in the scan frame with on-axis tip position (\textcircled{\raisebox{-1pt}{3}}). We conclude that the influence of the changed potential landscape must be smaller than the experimental accuracy. 
Furthermore, finite bias measurements confirmed that the energy gaps are independent of tip position. 
Therefore our experiments indicate that the activation energy of a confined $\nu=1/3$ state depends only weakly on the shape of the confining potential.

\begin{acknowledgments}
We thank R. Steinacher and C. R\"ossler for fruitfull discussions. The authors acknowledge financial support from ETH Z\"urich and
from the Swiss National Science Foundation (Schweizerischer Nationalfonds, NCCR "Quantum Science and Technology").
\end{acknowledgments}

\bibliography{plateaubib2}

\begin{thebibliography}{28}%
\makeatletter
\providecommand \@ifxundefined [1]{%
 \@ifx{#1\undefined}
}%
\providecommand \@ifnum [1]{%
 \ifnum #1\expandafter \@firstoftwo
 \else \expandafter \@secondoftwo
 \fi
}%
\providecommand \@ifx [1]{%
 \ifx #1\expandafter \@firstoftwo
 \else \expandafter \@secondoftwo
 \fi
}%
\providecommand \natexlab [1]{#1}%
\providecommand \enquote  [1]{``#1''}%
\providecommand \bibnamefont  [1]{#1}%
\providecommand \bibfnamefont [1]{#1}%
\providecommand \citenamefont [1]{#1}%
\providecommand \href@noop [0]{\@secondoftwo}%
\providecommand \href [0]{\begingroup \@sanitize@url \@href}%
\providecommand \@href[1]{\@@startlink{#1}\@@href}%
\providecommand \@@href[1]{\endgroup#1\@@endlink}%
\providecommand \@sanitize@url [0]{\catcode `\\12\catcode `\$12\catcode
  `\&12\catcode `\#12\catcode `\^12\catcode `\_12\catcode `\%12\relax}%
\providecommand \@@startlink[1]{}%
\providecommand \@@endlink[0]{}%
\providecommand \url  [0]{\begingroup\@sanitize@url \@url }%
\providecommand \@url [1]{\endgroup\@href {#1}{\urlprefix }}%
\providecommand \urlprefix  [0]{URL }%
\providecommand \Eprint [0]{\href }%
\providecommand \doibase [0]{http://dx.doi.org/}%
\providecommand \selectlanguage [0]{\@gobble}%
\providecommand \bibinfo  [0]{\@secondoftwo}%
\providecommand \bibfield  [0]{\@secondoftwo}%
\providecommand \translation [1]{[#1]}%
\providecommand \BibitemOpen [0]{}%
\providecommand \bibitemStop [0]{}%
\providecommand \bibitemNoStop [0]{.\EOS\space}%
\providecommand \EOS [0]{\spacefactor3000\relax}%
\providecommand \BibitemShut  [1]{\csname bibitem#1\endcsname}%
\let\auto@bib@innerbib\@empty
\bibitem [{\citenamefont {Klitzing}\ \emph {et~al.}(1980)\citenamefont
  {Klitzing}, \citenamefont {Dorda},\ and\ \citenamefont
  {Pepper}}]{klitzing_new_1980}%
  \BibitemOpen
  \bibfield  {author} {\bibinfo {author} {\bibfnamefont {K.~v.}\ \bibnamefont
  {Klitzing}}, \bibinfo {author} {\bibfnamefont {G.}~\bibnamefont {Dorda}}, \
  and\ \bibinfo {author} {\bibfnamefont {M.}~\bibnamefont {Pepper}},\ }\href
  {\doibase 10.1103/PhysRevLett.45.494} {\bibfield  {journal} {\bibinfo
  {journal} {Phys. Rev. Lett.}\ }\textbf {\bibinfo {volume} {45}},\ \bibinfo
  {pages} {494} (\bibinfo {year} {1980})}\BibitemShut {NoStop}%
\bibitem [{\citenamefont {Tsui}\ \emph {et~al.}(1982)\citenamefont {Tsui},
  \citenamefont {Stormer},\ and\ \citenamefont
  {Gossard}}]{tsui_two-dimensional_1982}%
  \BibitemOpen
  \bibfield  {author} {\bibinfo {author} {\bibfnamefont {D.~C.}\ \bibnamefont
  {Tsui}}, \bibinfo {author} {\bibfnamefont {H.~L.}\ \bibnamefont {Stormer}}, \
  and\ \bibinfo {author} {\bibfnamefont {A.~C.}\ \bibnamefont {Gossard}},\
  }\href {\doibase 10.1103/PhysRevLett.48.1559} {\bibfield  {journal} {\bibinfo
   {journal} {Phys. Rev. Lett.}\ }\textbf {\bibinfo {volume} {48}},\ \bibinfo
  {pages} {1559} (\bibinfo {year} {1982})}\BibitemShut {NoStop}%
\bibitem [{\citenamefont {Chamon}\ \emph {et~al.}(1997)\citenamefont {Chamon},
  \citenamefont {Freed}, \citenamefont {Kivelson}, \citenamefont {Sondhi},\
  and\ \citenamefont {Wen}}]{de_c._chamon_two_1997}%
  \BibitemOpen
  \bibfield  {author} {\bibinfo {author} {\bibfnamefont {C.~d.~C.}\
  \bibnamefont {Chamon}}, \bibinfo {author} {\bibfnamefont {D.~E.}\
  \bibnamefont {Freed}}, \bibinfo {author} {\bibfnamefont {S.~A.}\ \bibnamefont
  {Kivelson}}, \bibinfo {author} {\bibfnamefont {S.~L.}\ \bibnamefont
  {Sondhi}}, \ and\ \bibinfo {author} {\bibfnamefont {X.~G.}\ \bibnamefont
  {Wen}},\ }\href {\doibase 10.1103/PhysRevB.55.2331} {\bibfield  {journal}
  {\bibinfo  {journal} {Phys. Rev. B}\ }\textbf {\bibinfo {volume} {55}},\
  \bibinfo {pages} {2331} (\bibinfo {year} {1997})}\BibitemShut {NoStop}%
\bibitem [{\citenamefont {Radu}\ \emph {et~al.}(2008)\citenamefont {Radu},
  \citenamefont {Miller}, \citenamefont {Marcus}, \citenamefont {Kastner},
  \citenamefont {Pfeiffer},\ and\ \citenamefont
  {West}}]{radu_quasi-particle_2008}%
  \BibitemOpen
  \bibfield  {author} {\bibinfo {author} {\bibfnamefont {I.~P.}\ \bibnamefont
  {Radu}}, \bibinfo {author} {\bibfnamefont {J.~B.}\ \bibnamefont {Miller}},
  \bibinfo {author} {\bibfnamefont {C.~M.}\ \bibnamefont {Marcus}}, \bibinfo
  {author} {\bibfnamefont {M.~A.}\ \bibnamefont {Kastner}}, \bibinfo {author}
  {\bibfnamefont {L.~N.}\ \bibnamefont {Pfeiffer}}, \ and\ \bibinfo {author}
  {\bibfnamefont {K.~W.}\ \bibnamefont {West}},\ }\href {\doibase
  10.1126/science.1157560} {\bibfield  {journal} {\bibinfo  {journal}
  {Science}\ }\textbf {\bibinfo {volume} {320}},\ \bibinfo {pages} {899}
  (\bibinfo {year} {2008})}\BibitemShut {NoStop}%
\bibitem [{\citenamefont {Lin}\ \emph {et~al.}(2012)\citenamefont {Lin},
  \citenamefont {Dillard}, \citenamefont {Kastner}, \citenamefont {Pfeiffer},\
  and\ \citenamefont {West}}]{lin_measurements_2012}%
  \BibitemOpen
  \bibfield  {author} {\bibinfo {author} {\bibfnamefont {X.}~\bibnamefont
  {Lin}}, \bibinfo {author} {\bibfnamefont {C.}~\bibnamefont {Dillard}},
  \bibinfo {author} {\bibfnamefont {M.~A.}\ \bibnamefont {Kastner}}, \bibinfo
  {author} {\bibfnamefont {L.~N.}\ \bibnamefont {Pfeiffer}}, \ and\ \bibinfo
  {author} {\bibfnamefont {K.~W.}\ \bibnamefont {West}},\ }\href {\doibase
  10.1103/PhysRevB.85.165321} {\bibfield  {journal} {\bibinfo  {journal} {Phys.
  Rev. B}\ }\textbf {\bibinfo {volume} {85}},\ \bibinfo {pages} {165321}
  (\bibinfo {year} {2012})}\BibitemShut {NoStop}%
\bibitem [{\citenamefont {Baer}\ \emph
  {et~al.}(2014{\natexlab{a}})\citenamefont {Baer}, \citenamefont {R\"ossler},
  \citenamefont {Ihn}, \citenamefont {Ensslin}, \citenamefont {Reichl},\ and\
  \citenamefont {Wegscheider}}]{baer_experimental_2014}%
  \BibitemOpen
  \bibfield  {author} {\bibinfo {author} {\bibfnamefont {S.}~\bibnamefont
  {Baer}}, \bibinfo {author} {\bibfnamefont {C.}~\bibnamefont {R\"ossler}},
  \bibinfo {author} {\bibfnamefont {T.}~\bibnamefont {Ihn}}, \bibinfo {author}
  {\bibfnamefont {K.}~\bibnamefont {Ensslin}}, \bibinfo {author} {\bibfnamefont
  {C.}~\bibnamefont {Reichl}}, \ and\ \bibinfo {author} {\bibfnamefont
  {W.}~\bibnamefont {Wegscheider}},\ }\href@noop {} {\bibfield  {journal}
  {\bibinfo  {journal} {Phys. Rev. B}\ }\textbf {\bibinfo {volume} {90}},\
  \bibinfo {pages} {075403} (\bibinfo {year} {2014}{\natexlab{a}})}\BibitemShut
  {NoStop}%
\bibitem [{\citenamefont {Camino}\ \emph {et~al.}(2005)\citenamefont {Camino},
  \citenamefont {Zhou},\ and\ \citenamefont
  {Goldman}}]{camino_realization_2005}%
  \BibitemOpen
  \bibfield  {author} {\bibinfo {author} {\bibfnamefont {F.~E.}\ \bibnamefont
  {Camino}}, \bibinfo {author} {\bibfnamefont {W.}~\bibnamefont {Zhou}}, \ and\
  \bibinfo {author} {\bibfnamefont {V.~J.}\ \bibnamefont {Goldman}},\ }\href
  {\doibase 10.1103/PhysRevB.72.075342} {\bibfield  {journal} {\bibinfo
  {journal} {Phys. Rev. B}\ }\textbf {\bibinfo {volume} {72}},\ \bibinfo
  {pages} {075342} (\bibinfo {year} {2005})}\BibitemShut {NoStop}%
\bibitem [{\citenamefont {Camino}\ \emph {et~al.}(2007)\citenamefont {Camino},
  \citenamefont {Zhou},\ and\ \citenamefont {Goldman}}]{camino_$e/3$_2007}%
  \BibitemOpen
  \bibfield  {author} {\bibinfo {author} {\bibfnamefont {F.~E.}\ \bibnamefont
  {Camino}}, \bibinfo {author} {\bibfnamefont {W.}~\bibnamefont {Zhou}}, \ and\
  \bibinfo {author} {\bibfnamefont {V.~J.}\ \bibnamefont {Goldman}},\ }\href
  {\doibase 10.1103/PhysRevLett.98.076805} {\bibfield  {journal} {\bibinfo
  {journal} {Phys. Rev. Lett.}\ }\textbf {\bibinfo {volume} {98}},\ \bibinfo
  {pages} {076805} (\bibinfo {year} {2007})}\BibitemShut {NoStop}%
\bibitem [{\citenamefont {Kou}\ \emph {et~al.}(2012)\citenamefont {Kou},
  \citenamefont {Marcus}, \citenamefont {Pfeiffer},\ and\ \citenamefont
  {West}}]{kou_coulomb_2012}%
  \BibitemOpen
  \bibfield  {author} {\bibinfo {author} {\bibfnamefont {A.}~\bibnamefont
  {Kou}}, \bibinfo {author} {\bibfnamefont {C.~M.}\ \bibnamefont {Marcus}},
  \bibinfo {author} {\bibfnamefont {L.~N.}\ \bibnamefont {Pfeiffer}}, \ and\
  \bibinfo {author} {\bibfnamefont {K.~W.}\ \bibnamefont {West}},\ }\href
  {\doibase 10.1103/PhysRevLett.108.256803} {\bibfield  {journal} {\bibinfo
  {journal} {Phys. Rev. Lett.}\ }\textbf {\bibinfo {volume} {108}},\ \bibinfo
  {pages} {256803} (\bibinfo {year} {2012})}\BibitemShut {NoStop}%
\bibitem [{\citenamefont {Willett}\ \emph {et~al.}(2013)\citenamefont
  {Willett}, \citenamefont {Nayak}, \citenamefont {Shtengel}, \citenamefont
  {Pfeiffer},\ and\ \citenamefont {West}}]{willett_magnetic-field-tuned_2013}%
  \BibitemOpen
  \bibfield  {author} {\bibinfo {author} {\bibfnamefont {R.~L.}\ \bibnamefont
  {Willett}}, \bibinfo {author} {\bibfnamefont {C.}~\bibnamefont {Nayak}},
  \bibinfo {author} {\bibfnamefont {K.}~\bibnamefont {Shtengel}}, \bibinfo
  {author} {\bibfnamefont {L.~N.}\ \bibnamefont {Pfeiffer}}, \ and\ \bibinfo
  {author} {\bibfnamefont {K.~W.}\ \bibnamefont {West}},\ }\href {\doibase
  10.1103/PhysRevLett.111.186401} {\bibfield  {journal} {\bibinfo  {journal}
  {Phys. Rev. Lett.}\ }\textbf {\bibinfo {volume} {111}},\ \bibinfo {pages}
  {186401} (\bibinfo {year} {2013})}\BibitemShut {NoStop}%
\bibitem [{\citenamefont {Aoki}\ \emph {et~al.}(2005)\citenamefont {Aoki},
  \citenamefont {da~Cunha}, \citenamefont {Akis}, \citenamefont {Ferry},\ and\
  \citenamefont {Ochiai}}]{aoki_imaging_2005}%
  \BibitemOpen
  \bibfield  {author} {\bibinfo {author} {\bibfnamefont {N.}~\bibnamefont
  {Aoki}}, \bibinfo {author} {\bibfnamefont {C.~R.}\ \bibnamefont {da~Cunha}},
  \bibinfo {author} {\bibfnamefont {R.}~\bibnamefont {Akis}}, \bibinfo {author}
  {\bibfnamefont {D.~K.}\ \bibnamefont {Ferry}}, \ and\ \bibinfo {author}
  {\bibfnamefont {Y.}~\bibnamefont {Ochiai}},\ }\href {\doibase
  10.1103/PhysRevB.72.155327} {\bibfield  {journal} {\bibinfo  {journal} {Phys.
  Rev. B}\ }\textbf {\bibinfo {volume} {72}},\ \bibinfo {pages} {155327}
  (\bibinfo {year} {2005})}\BibitemShut {NoStop}%
\bibitem [{\citenamefont {Paradiso}\ \emph {et~al.}(2012)\citenamefont
  {Paradiso}, \citenamefont {Heun}, \citenamefont {Roddaro}, \citenamefont
  {Sorba}, \citenamefont {Beltram}, \citenamefont {Biasiol}, \citenamefont
  {Pfeiffer},\ and\ \citenamefont {West}}]{paradiso_imaging_2012}%
  \BibitemOpen
  \bibfield  {author} {\bibinfo {author} {\bibfnamefont {N.}~\bibnamefont
  {Paradiso}}, \bibinfo {author} {\bibfnamefont {S.}~\bibnamefont {Heun}},
  \bibinfo {author} {\bibfnamefont {S.}~\bibnamefont {Roddaro}}, \bibinfo
  {author} {\bibfnamefont {L.}~\bibnamefont {Sorba}}, \bibinfo {author}
  {\bibfnamefont {F.}~\bibnamefont {Beltram}}, \bibinfo {author} {\bibfnamefont
  {G.}~\bibnamefont {Biasiol}}, \bibinfo {author} {\bibfnamefont {L.~N.}\
  \bibnamefont {Pfeiffer}}, \ and\ \bibinfo {author} {\bibfnamefont {K.~W.}\
  \bibnamefont {West}},\ }\href {\doibase 10.1103/PhysRevLett.108.246801}
  {\bibfield  {journal} {\bibinfo  {journal} {Phys. Rev. Lett.}\ }\textbf
  {\bibinfo {volume} {108}},\ \bibinfo {pages} {246801} (\bibinfo {year}
  {2012})}\BibitemShut {NoStop}%
\bibitem [{\citenamefont {Pascher}\ \emph
  {et~al.}(2014{\natexlab{a}})\citenamefont {Pascher}, \citenamefont
  {R\"ossler}, \citenamefont {Ihn}, \citenamefont {Ensslin}, \citenamefont
  {Reichl},\ and\ \citenamefont {Wegscheider}}]{pascher_imaging_2014}%
  \BibitemOpen
  \bibfield  {author} {\bibinfo {author} {\bibfnamefont {N.}~\bibnamefont
  {Pascher}}, \bibinfo {author} {\bibfnamefont {C.}~\bibnamefont {R\"ossler}},
  \bibinfo {author} {\bibfnamefont {T.}~\bibnamefont {Ihn}}, \bibinfo {author}
  {\bibfnamefont {K.}~\bibnamefont {Ensslin}}, \bibinfo {author} {\bibfnamefont
  {C.}~\bibnamefont {Reichl}}, \ and\ \bibinfo {author} {\bibfnamefont
  {W.}~\bibnamefont {Wegscheider}},\ }\href {\doibase
  10.1103/PhysRevX.4.011014} {\bibfield  {journal} {\bibinfo  {journal} {Phys.
  Rev. X}\ }\textbf {\bibinfo {volume} {4}},\ \bibinfo {pages} {011014}
  (\bibinfo {year} {2014}{\natexlab{a}})}\BibitemShut {NoStop}%
\bibitem [{\citenamefont {Baer}\ \emph
  {et~al.}(2014{\natexlab{b}})\citenamefont {Baer}, \citenamefont {{C.
  R\"ossler}}, \citenamefont {de~Wiljes}, \citenamefont {Ardelt}, \citenamefont
  {Ihn}, \citenamefont {Ensslin}, \citenamefont {Reichl},\ and\ \citenamefont
  {Wegscheider}}]{baer_interplay_2014}%
  \BibitemOpen
  \bibfield  {author} {\bibinfo {author} {\bibfnamefont {S.}~\bibnamefont
  {Baer}}, \bibinfo {author} {\bibnamefont {{C. R\"ossler}}}, \bibinfo {author}
  {\bibfnamefont {E.~C.}\ \bibnamefont {de~Wiljes}}, \bibinfo {author}
  {\bibfnamefont {P.-L.}\ \bibnamefont {Ardelt}}, \bibinfo {author}
  {\bibfnamefont {T.}~\bibnamefont {Ihn}}, \bibinfo {author} {\bibfnamefont
  {K.}~\bibnamefont {Ensslin}}, \bibinfo {author} {\bibfnamefont
  {C.}~\bibnamefont {Reichl}}, \ and\ \bibinfo {author} {\bibfnamefont
  {W.}~\bibnamefont {Wegscheider}},\ }\href@noop {} {\bibfield  {journal}
  {\bibinfo  {journal} {Phys. Rev. B}\ }\textbf {\bibinfo {volume} {89}},\
  \bibinfo {pages} {085424} (\bibinfo {year} {2014}{\natexlab{b}})}\BibitemShut
  {NoStop}%
\bibitem [{\citenamefont {Dethlefsen}\ \emph {et~al.}(2006)\citenamefont
  {Dethlefsen}, \citenamefont {Mariani}, \citenamefont {Tranitz}, \citenamefont
  {Wegscheider},\ and\ \citenamefont {Haug}}]{dethlefsen_signatures_2006}%
  \BibitemOpen
  \bibfield  {author} {\bibinfo {author} {\bibfnamefont {A.~F.}\ \bibnamefont
  {Dethlefsen}}, \bibinfo {author} {\bibfnamefont {E.}~\bibnamefont {Mariani}},
  \bibinfo {author} {\bibfnamefont {H.-P.}\ \bibnamefont {Tranitz}}, \bibinfo
  {author} {\bibfnamefont {W.}~\bibnamefont {Wegscheider}}, \ and\ \bibinfo
  {author} {\bibfnamefont {R.~J.}\ \bibnamefont {Haug}},\ }\href {\doibase
  10.1103/PhysRevB.74.165325} {\bibfield  {journal} {\bibinfo  {journal} {Phys.
  Rev. B}\ }\textbf {\bibinfo {volume} {74}},\ \bibinfo {pages} {165325}
  (\bibinfo {year} {2006})}\BibitemShut {NoStop}%
\bibitem [{\citenamefont {Gildemeister}\ \emph {et~al.}(2007)\citenamefont
  {Gildemeister}, \citenamefont {Ihn}, \citenamefont {Barengo}, \citenamefont
  {Studerus},\ and\ \citenamefont {Ensslin}}]{gildemeister_construction_2007}%
  \BibitemOpen
  \bibfield  {author} {\bibinfo {author} {\bibfnamefont {A.~E.}\ \bibnamefont
  {Gildemeister}}, \bibinfo {author} {\bibfnamefont {T.}~\bibnamefont {Ihn}},
  \bibinfo {author} {\bibfnamefont {C.}~\bibnamefont {Barengo}}, \bibinfo
  {author} {\bibfnamefont {P.}~\bibnamefont {Studerus}}, \ and\ \bibinfo
  {author} {\bibfnamefont {K.}~\bibnamefont {Ensslin}},\ }\href {\doibase
  10.1063/1.2431793} {\bibfield  {journal} {\bibinfo  {journal} {Rev Sci
  Instrum}\ }\textbf {\bibinfo {volume} {78}},\ \bibinfo {pages} {013704}
  (\bibinfo {year} {2007})}\BibitemShut {NoStop}%
\bibitem [{\citenamefont {Rychen}\ \emph {et~al.}(1999)\citenamefont {Rychen},
  \citenamefont {Ihn}, \citenamefont {Studerus}, \citenamefont {Herrmann},\
  and\ \citenamefont {Ensslin}}]{rychen_low-temperature_1999}%
  \BibitemOpen
  \bibfield  {author} {\bibinfo {author} {\bibfnamefont {J.}~\bibnamefont
  {Rychen}}, \bibinfo {author} {\bibfnamefont {T.}~\bibnamefont {Ihn}},
  \bibinfo {author} {\bibfnamefont {P.}~\bibnamefont {Studerus}}, \bibinfo
  {author} {\bibfnamefont {A.}~\bibnamefont {Herrmann}}, \ and\ \bibinfo
  {author} {\bibfnamefont {K.}~\bibnamefont {Ensslin}},\ }\href {\doibase
  10.1063/1.1149842} {\bibfield  {journal} {\bibinfo  {journal} {Rev Sci
  Instrum}\ }\textbf {\bibinfo {volume} {70}},\ \bibinfo {pages} {2765}
  (\bibinfo {year} {1999})}\BibitemShut {NoStop}%
\bibitem [{\citenamefont {Beenakker}\ and\ \citenamefont {van
  Houten}(1991)}]{beenakker_quantum_1991}%
  \BibitemOpen
  \bibfield  {author} {\bibinfo {author} {\bibfnamefont {C.~W.~J.}\
  \bibnamefont {Beenakker}}\ and\ \bibinfo {author} {\bibfnamefont
  {H.}~\bibnamefont {van Houten}},\ }\href {\doibase
  10.1016/S0081-1947(08)60091-0} {\bibfield  {journal} {\bibinfo  {journal}
  {Solid State Physics}\ }\textbf {\bibinfo {volume} {44}},\ \bibinfo {pages}
  {1} (\bibinfo {year} {1991})},\ \bibinfo {note} {solid State Physics 44, 1
  (1991)}\BibitemShut {NoStop}%
\bibitem [{Note1()}]{Note1}%
  \BibitemOpen
  \bibinfo {note} {The tip can not be moved over the upper top gate because of
  an obstacle that would cause a tip crash.}\BibitemShut {Stop}%
\bibitem [{\citenamefont {Buttiker}(1990)}]{buettiker_quantized_1990}%
  \BibitemOpen
  \bibfield  {author} {\bibinfo {author} {\bibfnamefont {M.}~\bibnamefont
  {Buttiker}},\ }\href {\doibase 10.1103/PhysRevB.41.7906} {\bibfield
  {journal} {\bibinfo  {journal} {Phys. Rev. B}\ }\textbf {\bibinfo {volume}
  {41}},\ \bibinfo {pages} {7906} (\bibinfo {year} {1990})}\BibitemShut
  {NoStop}%
\bibitem [{\citenamefont {Pascher}\ \emph
  {et~al.}(2014{\natexlab{b}})\citenamefont {Pascher}, \citenamefont {Timpu},
  \citenamefont {R\"ossler}, \citenamefont {Ihn}, \citenamefont {Ensslin},
  \citenamefont {Reichl},\ and\ \citenamefont
  {Wegscheider}}]{pascher_resonant_2014-1}%
  \BibitemOpen
  \bibfield  {author} {\bibinfo {author} {\bibfnamefont {N.}~\bibnamefont
  {Pascher}}, \bibinfo {author} {\bibfnamefont {F.}~\bibnamefont {Timpu}},
  \bibinfo {author} {\bibfnamefont {C.}~\bibnamefont {R\"ossler}}, \bibinfo
  {author} {\bibfnamefont {T.}~\bibnamefont {Ihn}}, \bibinfo {author}
  {\bibfnamefont {K.}~\bibnamefont {Ensslin}}, \bibinfo {author} {\bibfnamefont
  {C.}~\bibnamefont {Reichl}}, \ and\ \bibinfo {author} {\bibfnamefont
  {W.}~\bibnamefont {Wegscheider}},\ }\href {\doibase
  10.1103/PhysRevB.89.245408} {\bibfield  {journal} {\bibinfo  {journal} {Phys.
  Rev. B}\ }\textbf {\bibinfo {volume} {89}},\ \bibinfo {pages} {245408}
  (\bibinfo {year} {2014}{\natexlab{b}})}\BibitemShut {NoStop}%
\bibitem [{Note2()}]{Note2}%
  \BibitemOpen
  \bibinfo {note} {Electrostatic study using \protect \emph {COMSOL
  5.0}}\BibitemShut {NoStop}%
\bibitem [{\citenamefont {Ando}\ \emph {et~al.}(1982)\citenamefont {Ando},
  \citenamefont {Fowler},\ and\ \citenamefont {Stern}}]{ando_electronic_1982}%
  \BibitemOpen
  \bibfield  {author} {\bibinfo {author} {\bibfnamefont {T.}~\bibnamefont
  {Ando}}, \bibinfo {author} {\bibfnamefont {A.~B.}\ \bibnamefont {Fowler}}, \
  and\ \bibinfo {author} {\bibfnamefont {F.}~\bibnamefont {Stern}},\ }\href
  {\doibase 10.1103/RevModPhys.54.437} {\bibfield  {journal} {\bibinfo
  {journal} {Rev. Mod. Phys.}\ }\textbf {\bibinfo {volume} {54}},\ \bibinfo
  {pages} {437} (\bibinfo {year} {1982})}\BibitemShut {NoStop}%
\bibitem [{\citenamefont {Fang}\ and\ \citenamefont
  {Howard}(1966)}]{fang_negative_1966}%
  \BibitemOpen
  \bibfield  {author} {\bibinfo {author} {\bibfnamefont {F.~F.}\ \bibnamefont
  {Fang}}\ and\ \bibinfo {author} {\bibfnamefont {W.~E.}\ \bibnamefont
  {Howard}},\ }\href {\doibase 10.1103/PhysRevLett.16.797} {\bibfield
  {journal} {\bibinfo  {journal} {Phys. Rev. Lett.}\ }\textbf {\bibinfo
  {volume} {16}},\ \bibinfo {pages} {797} (\bibinfo {year} {1966})}\BibitemShut
  {NoStop}%
\bibitem [{Note3()}]{Note3}%
  \BibitemOpen
  \bibinfo {note} {$ | \nabla I_{\protect \mathrm {SD}}|=\protect \sqrt
  {(\protect \mathrm {d}I_{\protect \mathrm {SD}}/\protect \mathrm
  {d}x)^2+(\protect \mathrm {d}I_{\protect \mathrm {SD}}/\protect \mathrm
  {d}y)^2}$}\BibitemShut {NoStop}%
\bibitem [{\citenamefont {Wei}\ \emph {et~al.}(1985)\citenamefont {Wei},
  \citenamefont {Chang}, \citenamefont {Tsui},\ and\ \citenamefont
  {Razeghi}}]{wei_temperature_1985}%
  \BibitemOpen
  \bibfield  {author} {\bibinfo {author} {\bibfnamefont {H.~P.}\ \bibnamefont
  {Wei}}, \bibinfo {author} {\bibfnamefont {A.~M.}\ \bibnamefont {Chang}},
  \bibinfo {author} {\bibfnamefont {D.~C.}\ \bibnamefont {Tsui}}, \ and\
  \bibinfo {author} {\bibfnamefont {M.}~\bibnamefont {Razeghi}},\ }\href
  {\doibase 10.1103/PhysRevB.32.7016} {\bibfield  {journal} {\bibinfo
  {journal} {Phys. Rev. B}\ }\textbf {\bibinfo {volume} {32}},\ \bibinfo
  {pages} {7016} (\bibinfo {year} {1985})}\BibitemShut {NoStop}%
\bibitem [{\citenamefont {Ebert}\ \emph {et~al.}(1983)\citenamefont {Ebert},
  \citenamefont {von Klitzing}, \citenamefont {Ploog},\ and\ \citenamefont
  {Weimann}}]{ebert_two-dimensional_1983}%
  \BibitemOpen
  \bibfield  {author} {\bibinfo {author} {\bibfnamefont {M.}~\bibnamefont
  {Ebert}}, \bibinfo {author} {\bibfnamefont {K.}~\bibnamefont {von Klitzing}},
  \bibinfo {author} {\bibfnamefont {M.}~\bibnamefont {Ploog}}, \ and\ \bibinfo
  {author} {\bibfnamefont {G.}~\bibnamefont {Weimann}},\ }\href@noop {}
  {\bibfield  {journal} {\bibinfo  {journal} {J. Phys. C}\ ,\ \bibinfo {pages}
  {5441}} (\bibinfo {year} {1983})}\BibitemShut {NoStop}%
\bibitem [{\citenamefont {Berggren}\ \emph {et~al.}(1988)\citenamefont
  {Berggren}, \citenamefont {Roos},\ and\ \citenamefont {van
  Houten}}]{berggren_characterization_1988}%
  \BibitemOpen
  \bibfield  {author} {\bibinfo {author} {\bibfnamefont {K.-F.}\ \bibnamefont
  {Berggren}}, \bibinfo {author} {\bibfnamefont {G.}~\bibnamefont {Roos}}, \
  and\ \bibinfo {author} {\bibfnamefont {H.}~\bibnamefont {van Houten}},\
  }\href {\doibase 10.1103/PhysRevB.37.10118} {\bibfield  {journal} {\bibinfo
  {journal} {Phys. Rev. B}\ }\textbf {\bibinfo {volume} {37}},\ \bibinfo
  {pages} {10118} (\bibinfo {year} {1988})}\BibitemShut {NoStop}%
\end{thebibliography}%

\end{document}